\documentclass[aps,pra,preprint,showpacs]{revtex4}
\usepackage{graphicx}
\usepackage{epsfig}

\begin{document}

\title{Binary Quantum Search}

\author{Vladimir E. Korepin}
\email{korepin@insti.physics.sunysb.edu}
\affiliation{C.N. Yang Institute for Theoretical Physics,
 State University of New York at Stony Brook,\\
 Stony Brook, NY 11794-3840}
 
\author{Ying Xu}
\email{yixu@ic.sunysb.edu}
\affiliation{Department of Physics and Astronomy,
 State University of New York at Stony Brook,\\
 Stony Brook, NY 11794-3800}

\date{\today}

\begin{abstract}
Database search has wide applications and is used as a subroutine in many important algorithms. We shall consider a database with one target item. Quantum algorithm finds the target item in a database faster than any classical algorithm. It frequently occurs in practice that only a portion of information about the target item is interesting, or we need to find a group of items sharing some common feature as the target item. This problem is in general formulated as search for a part of the database [a block] containing the target item, instead of the item itself. This is partial search. Partial search trades accuracy for speed, i.e. it works faster than a full search. Partial search algorithm was discovered by Grover and Radhakrishnan. We shall consider optimized version of the algorithm and call it GRK. It can be applied successively [in a sequence]. First the database is partitioned into blocks and we use GRK to find the target block. Then this target block is partitioned into sub-blocks and we use GRK again to find the target sub-block. [We can call it binary quantum search.] Another possibility is to partition the database into sub-blocks directly and use GRK to find the target sub-block  in one time. In this paper we prove that the latter is faster [makes less queries to the oracle].
\end{abstract}


\maketitle

\section{Introduction}
Database search has many applications. Search algorithm enters as a subroutine in many important algorithms \cite{CLLRS, NC, CEHMM} in computer sciences. Grover discovered a quantum algorithm \cite{Grover} which searches a database faster than any classical algorithm.
Let's consider a database with one target item. We use number of queries to the oracle as complexity measure. The Grover algorithm finds the target item [with probability 1] in 
\begin{equation}
j_{\mbox{\scriptsize{full}}}= \frac{\pi}{4} \sqrt{N}, \qquad N\rightarrow\infty \label{full}
\end{equation}
iterations [queries to the oracle]. We shall call it a full search. 

It occurs frequently in practice that less information is needed. For example, the address of the target item in binary form is $\left| t \right\rangle = \left|b_{1}b_{2}b_{3}...b_{n}\right\rangle$, and we want to find only the first 3 bits  $b_{1}b_{2}b_{3}$. This means that the database is partitioned into 8 blocks. All items in a block share the common feature such that the first 3 bits being the same. We want to find the block containing the target item. This is an example of partial search. The general problem of partial search considers the following: An $N$ item database is partitioned into $K$ blocks, each of the same size 
\begin{equation}
b=\frac{N}{K} \label{size}.
\end{equation}
A user wants to find the block containing the target item, instead of the target item itself. The block with the target item is called the target block; others non-target blocks. Partial search naturally arises in list matching \cite{Heiligman}. Partial search is not only a compromise on accuracy for speed, but also has it own significance. The GRK algorithm of partial search was suggested by Grover and Radhakrishnan \cite{jaik}, and optimized in \cite{ Korepin}. It takes \newline $\sim \frac{\pi}{4}(1-\mbox{coeff}(K))\sqrt{N}$ number of queries to find the target block. Here coeff$(K)$ is a finite positive number, which depends on $K$ and has a limit when blocks are large $b\rightarrow\infty$. GRK is the most efficient partial search algorithm known in literatures \cite{jaik, KL, CK, Korepin, KV}.

GRK can be applied in a sequence [one after another], i.e. after the first GRK, the target block found can be further partitioned into sub-blocks. Then a second GRK can be applied to find the sub-block containing the target item [called the target sub-block]. We shall call the sequence of GRK's a partial search hierarchy. In hierarchical search we iterate GRK. A practical example would be: In order to find a hotel, we first look at a State map and then a town map. We shall see that the second GRK works faster than the first one. Actually, GRK can be conducted repeatedly until we find the smallest target sub-sub-block interested. The total number of queries is the sum of queries of each GRK in the hierarchy. [We use number of queries as measure of complexity.] 

Alternative to a partial search hierarchy which finds the target sub-sub-block, we could partition the database directly into sub-sub-blocks and use GRK once: We shall call it direct partial search. Although each GRK works faster than the previous one in the hierarchy, it is not guaranteed that the total number of queries in the hierarchy [sequence of GRK's] is less than that of a direct partial search. On the contrary, we will prove that \textit{direct partial search works faster}, which is the main result of the paper. For example, consider a database partitioned into 2 blocks. Each block is partitioned into 2 sub-blocks, so totally 4 sub-blocks. One could first find the target block using GRK, then the target sub-block using sequential GRK. However, it is faster to run a GRK directly over the 4 sub-blocks, which finds the target sub-block once.

The  paper consists of two parts:  In the first part, we start with the Grover algorithm and the GRK algorithm. Then we study the partial search hierarchy in detail. The second part proceeds to a comparison of the hierarchical partial search with direct partial search. Then we prove our main result that direct partial search works faster.

\section{The Grover Search Algorithm}

	In our paper, we consider different methods of partial search. They are all built on the original idea of the full Grover search \cite{Grover, NC, Mosca}. Let's formulate the problem. Consider a database of $N$ items with one target item \footnote{Target item also called in literatures marked item or solution.}. The database is associated with a Hilbert space with $N$ normalized basis vectors. The basis vector corresponding to item $x$ is denoted by $|x\rangle$. The Grover search is a quantum algorithm which starts from the uniform superposition of all basis vectors in the whole database:
\begin{equation}
|s_1\rangle = \frac{1}{\sqrt{N}}\sum_{x=0}^{N-1}|x\rangle , \qquad 
\langle s_1|s_1\rangle =1 . \label{ave}
\end{equation}
The algorithm searches for a single target item $|t\rangle$ iteratively. The Grover iteration is a unitary transform:
\begin{equation}
G_1=-I_{s_1}I_t. \label{iter}
\end{equation}
Later we shall call it a global iteration in GRK. Here $I_t$ and $I_{s_1}$
are two inversions about the target item $|t\rangle$ and the uniform superposition $|s_1\rangle$ defined in (\ref{ave}), respectively:
\begin{eqnarray}
I_t=\hat{I}-2|t\rangle \langle t| \label{target}, \\
I_{s_1}=\hat{I}-2|s_1\rangle \langle s_1|, \label{average}
\end{eqnarray}
where $\hat{I}$ is the identical operator. The Grover iteration $G_1$ is a rotation \cite{Mosca} in the Hilbert space from $|s_1\rangle$ towards the target $|t\rangle$ by an angle $\theta_1$ defined by:
\begin{equation}
\sin ^2\theta_1 =\frac{1}{N}. \label{ang1}
\end{equation}
After $j_1$ iterations the state of the database becomes \cite{Brass, Mosca, NC}:
\begin{eqnarray}
G_1^{j_1} |s_1\rangle  = \sin \left( (2j_1+1)\theta_1 \right) |t\rangle + 
\frac{\cos \left( (2j_1+1)\theta_1 \right)}{\sqrt{N-1}}
\sum^{\stackrel{\mbox{\tiny{N-1}}}{\mbox{\tiny{items}}}}_{\mbox{\scriptsize{x $\neq$ t}}} |x\rangle.
\label{first}
\end{eqnarray} 
Therefore after $j_{\mbox{\scriptsize{full}}}={\pi}/({4\theta_1})-{1}/{2}$ iterations the probability amplitude of $|t\rangle$ becomes unity and amplitudes of other items all vanish. i.e.
\begin{eqnarray}
G_1^{j_{\mbox{\scriptsize{full}}}} |s_1\rangle = |t\rangle.
\end{eqnarray}
As $N$ becomes large $j_{\mbox{\scriptsize{full}}}={\pi}/({4\theta_1})-{1}/{2}$ approaches (\ref{full}). More details on Grover search can be found in \cite{NC}.

\section{Algorithms for Partial Search}

Before introducing the GRK partial search algorithm [see next section], it worth mentioning a few other algorithms for comparison:

\begin{description}

	\item[a)] Naive Search
	
	Pick a block randomly and make a full Grover search in it [which makes $\frac{\pi}{4} \sqrt{\frac{N}{K}}$ queries to the oracle]. If we find the target item then we understand that this is the target block. If not, then we discard this block and pick another randomly. Make a full Grover search in it and repeat this procedure till we find the target block. In the worst case the target block will be the last one. So with probability 1 we have to use 
\begin{eqnarray}
	r(N,K)= \frac{(K-1)}{\sqrt{K}}\frac{\pi}{4}\sqrt{N} \label{naive} 
\end{eqnarray}
iterations [queries] to find the target block, see \footnote{$(\sqrt{K}/2)(\pi/4)\sqrt{N}$ queries on average.}.
	
A full Grover search finds the target item in $(\pi/4)\sqrt{N}$ queries. If we know the exact address of the target item then we also know the target block. Comparing $(\pi/4)\sqrt{N}$ with $r(N,K)$ in (\ref{naive}), we see that the naive version is faster only for two blocks $K=2$. [If $K\geq3$ a full search is faster].

	\item[b)] Binary Search
	
	Assume that $K=2^k$ with $k$ a positive integer. Divide the database into two blocks and make a full Grover search in one block. If the target item not found, then take the remaining block and divide it into two sub-blocks. Pick a sub-block randomly and make a full search again in it. Repeat the procedure until we are left with the last block. In the worst case, the number of queries necessary to find the target block is
\begin{eqnarray}
\frac{\pi}{4}\sqrt{N}\left\{\frac{1}{\sqrt{2}}+\frac{1}{\sqrt{4}}+\ldots+\frac{1}{2^{k/2}}\right\}, \qquad k=\log_{2}K.  \label{binary}
\end{eqnarray}
		
	The first two terms in the braces of (\ref{binary}) are greater than 1 for $K\geq3$,  
\begin{eqnarray}
	\frac{1}{\sqrt{2}}+\frac{1}{2}=\frac{\sqrt{2}+1}{2}>1.
\end{eqnarray}
	So this algorithm is less efficient than a full Grover search, when $K>2$.

	\item[c)] Grover and Radhakrishnan Version
	
	 A faster version was found in \cite{jaik}. Pick randomly a block and
make a full Grover search in the compliment [all items in the rest of the database]. Either the target item [and block] is found after the search or the picked block is the target block. This requires $\frac{\pi}{4}\sqrt{b(K-1)}=\frac{\pi}{4}\sqrt{N}\sqrt{\frac{K-1}{K}}$ queries. It is faster than a full search.

\end{description}
	

\section{The GRK Partial Search Algorithm}
	
	Grover and Radhakrishnan also discovered a faster quantum algorithm \cite{jaik} for partial search, which uses the same oracle as the main Grover algorithm. [See Summary and Appendix D.] Partial search also starts from the uniform superposition of all basis states (\ref{ave}).
A general structure of the algorithm is \cite{jaik, KG, KL, KV, CK, Korepin}:

\begin{description}
	\item [Step 1.] Global iterations: $j_1$ standard Grover iterations (\ref{iter}). After this step the state of database is $G_1^{j_1}|s_1\rangle$.
	
	\item [Step 2.] Simultaneous local iterations in each block: $j_2$ local Grover iterations defined in (\ref{liter}) below. After step 2 the state of database is $G_2^{j_2}G_1^{j_1}|s_1\rangle$. \\
	Local iteration is defined by
	\begin{eqnarray}
		G_2=\bigoplus_{\mbox{\scriptsize blocks}} ^{K}G_2^{\stackrel{\mbox{\tiny{one}}}{\mbox{\tiny{block}}}}
	=-\left(\bigoplus_{\mbox{\scriptsize {blocks}}} ^{K}I_{s_2}\right)I_t. \label{liter}
	\end{eqnarray}
It is a direct sum of Grover iterations [called local queries] defined in each block
\begin{eqnarray}
	G_2^{\stackrel{\mbox{\tiny{one}}}{\mbox{\tiny{block}}}}=-I_{s_2}I_t . 
\end{eqnarray} 
In the expression $I_t$ is the same inversion (\ref{target}), i.e. query to the oracle. $I_{s_2}$ is a local inversion 
\begin{eqnarray}
	I_{s_2}=\hat{I}-2|s_2\rangle \langle s_2|.
\end{eqnarray}
Here $|s_2\rangle$ is the uniform superposition of items in one block
\begin{equation}
|s_2\rangle = \frac{1}{\sqrt{b}}\sum_{\mbox{\tiny{one block}}}^{\mbox{\tiny{b items}}}|x\rangle. \label{nblock}
\end{equation}
Local iteration $G_2$ is a the Grover iteration in each block done simultaneously in all blocks. $G_2$ acts trivially on non-target blocks. 
A non-trivial operation [rotation] is present only in the target block with new rotation angle $\theta_2$ defined by
\begin{equation}
\sin^2 \theta_2 =\frac{K}{N}=\frac{1}{b}. \label{ang2}
\end{equation}
Note that amplitudes of all  items in non-target blocks remain intact.
	
	\item [Step 3.] Location of the target block with a final global iteration \cite{KL, KV, Korepin}:\\
	We have to vanish amplitudes of all items in non-target blocks. We can do it by application of one more global iteration. The resulting state is
\begin{eqnarray}
	|d\rangle \equiv G_1G_2^{j_2}G_1^{j_1}|s_1\rangle = \sin \omega |t\rangle + \frac{\cos \omega}{\sqrt{b-1}} \sum^{\stackrel{\mbox{\tiny{b-1}}}{\mbox{\tiny{items}}}}_{\stackrel{x \neq t}{\mbox{\tiny{target block}}}}|x\rangle. \label{fin}
\end{eqnarray}
The final state (\ref{fin}) is expressed as a superposition over items in the target block only. This is realized by requiring that the amplitude of any non-target block vanishes after the partial search, i.e.
\begin{eqnarray}
	\langle x|d\rangle = 0. \label{vanish}
\end{eqnarray}
Here $x$ is an arbitrary item in any non-target block. This vanishing condition can be written explicitly as an equality for $j_1$ and $j_2$, see \cite{Korepin}. We shall call it a cancellation condition.
	
\end{description}

This partial search algorithm was further optimized in \cite{Korepin}. In the large block limit $b\rightarrow\infty$, the total number of items also large $N\rightarrow\infty$, while the ratio $K=N/b$ kept finite. Then the expression for rotation angles (\ref{ang1}) and (\ref{ang2}) simplifies
\begin{equation}
\theta_1\rightarrow\frac{1}{\sqrt{N}},\qquad \theta_2\rightarrow\frac{1}{\sqrt{b}}.\label{dva}
\end{equation}
It turns out convenient to rewrite numbers of iterations in a scale form \cite{jaik}
\begin{eqnarray}
j_1=\left(\frac{\pi}{4} -\frac {\eta}{\sqrt{K}}\right) \sqrt{N}, \qquad 
j_2=\frac {\alpha}{\sqrt{K}}\sqrt{N}. \label{para}	
\end{eqnarray}
Here $\eta$ and $\alpha$ are parameters of order 1 [they have a limit]. The ranges of these parameters are discussed in Appendix B. The vanishing condition (\ref{vanish}) in terms of these parameters simplifies in the large $b$ limit \cite{CK, Korepin}
\begin{equation}
\tan \left(\frac{2\eta}{\sqrt{K}} \right) =\frac{2\sqrt{K}\sin 2\alpha}{K
 -4\sin^2 \alpha }. \label{constraint}
\end{equation}
The total number of queries is
\begin{eqnarray}
	S(K)\equiv j_1+j_2+1 \stackrel{b\rightarrow\infty}{\longrightarrow} \left(\frac{\pi}{4}+\frac {\alpha -\eta}{\sqrt{K}} \right) \sqrt{N}. \label{number}
\end{eqnarray}
It was minimized [subject to the constraint (\ref{constraint})] in \cite{Korepin}. The minimum number of queries is achieved at
\begin{equation}
	\eta \left(K\right) = \frac{1}{2}\sqrt{K} \arctan \left(\frac{\sqrt{3K-4}}{K-2}\right), \qquad    
	\alpha \left(K\right) = \frac{1}{2} \arccos \left(\frac{K-2}{2(K-1)}\right) . \label{min}
\end{equation}
Thus the minimized number of queries of GRK partial search [as a function of $K$] is
\begin{eqnarray}
	S\left(K\right)\stackrel{b\rightarrow\infty}{\longrightarrow}\left(\frac{\pi}{4}+\frac {\alpha \left(K\right)-\eta\left(K\right)}{\sqrt{K}} \right) \sqrt{N}. \label{no.par}
\end{eqnarray}
A proof of (\ref{min}) being the minimum is given in Appendix C. Note that $\alpha-\eta$ is negative and number of blocks $K\geq2$ in a non-trivial situation.

In the large block limit, the $\omega$ appeared in (\ref{fin}) is
\begin{eqnarray}
	\omega=\alpha(K), \label{omega}
\end{eqnarray}
see \cite{Korepin}. As a consequence, the state of database after GRK (\ref{fin}) is the following: The amplitudes of items in non-target blocks all vanish and the state of the target block is
\begin{eqnarray}
	|d\rangle = \sin \alpha(K)|t\rangle + \frac{\cos \alpha(K)}{\sqrt{b-1}} \sum^{\stackrel{\mbox{\tiny{b-1}}}{\mbox{\tiny{items}}}}_{\stackrel{x \neq t}{\mbox{\tiny{target block}}}}|x\rangle. \label{final}
\end{eqnarray}


\section{The Partial Search Hierarchy}

A partial search hierarchy is a sequence of GRK's. After location of the target block, we may consider a subsequent GRK partial search: The target block is further partitioned into $\tilde{K}$ sub-blocks and we search for the sub-block containing the target item [target sub-block]. For example we can use Google Earth to find the State of New York first on the map of USA and then make a sequential search for Stony Brook in the State map. 

We shall show below that a sequential GRK can be done faster than the first GRK. The coefficient $\pi/4$  in (\ref{no.par}) is replaced by a smaller number:
\begin{equation}
\frac{\pi}{4} \rightarrow \frac{\pi}{4}- \frac{1}{4} \arccos \left( \frac{K-2}{2(K-1)}\right).\label{renorm}
\end{equation}
Each successive GRK works faster than the previous one for two reasons. First, the new database is smaller [only one block of the previous one]. Second, the initial state of the new database (\ref{final}) can be represented in different forms (\ref{ini}) and (\ref{ininew}) below. We see that for sequential GRK, the initial state is no longer a uniform superposition of basis vectors of the new database. It is an unevenly weighted superposition with emphasis on the target $|t\rangle$, see (\ref{ini}) and (\ref{ininew}). In other words, the new initial state of the database is equivalent to a partially searched [though not fully searched] one. This fact was studied in \cite{Korepin}. It was shown that after the first GRK the state of the target block [new database] can be written as [(\ref{final}) rewritten]
\begin{eqnarray}
	|d\rangle = G_1G_2^{j_2}G_1^{j_1}|s_1\rangle = \sin \alpha(K) |t\rangle + \frac{\cos \alpha(K)}{\sqrt{b-1}} \sum^{\stackrel{\mbox{\tiny{b-1}}}{\mbox{\tiny{items}}}}_{\stackrel{x \neq t}{\mbox{\tiny{target block}}}}|x\rangle. \label{state}
\end{eqnarray}
We have used relation (\ref{omega}). Compared with (\ref{first}), we see that the state after the first GRK (\ref{state}) takes the form
\begin{eqnarray}
	|d\rangle=G_1G_2^{j_2}G_1^{j_1}|s_1\rangle = G_2^{\frac{\alpha\left(K\right)}{2}\sqrt{b}}|s_2\rangle, \label{ini}
\end{eqnarray}
which serves as the initial state of the sequential GRK.

For notational convenience, we use a "$\sim$" to indicate variables in sequential GRK and make the following definitions:
\begin{eqnarray}
	\mbox{Number of items in new database}: \qquad \qquad \qquad \qquad
\tilde{N}=b=\tilde{K} \tilde{b}, \label{newnum}\\
	\mbox{Uniform superposition of new database}: \qquad \qquad \qquad \ \
|\tilde{s_1}\rangle = |s_2\rangle, \\
	\mbox{New global inversion}: \qquad \qquad \qquad \qquad \qquad \qquad \qquad \quad \ 
I_{\tilde{s}_1} = I_{s_2}, \\
	\mbox{New global iteration}: \qquad \qquad \qquad \qquad \qquad
\tilde{G_1} = G_2, \qquad \tilde{\theta_1} = \theta_2, \label{newang1} \\ \nonumber
	\\ \mbox{Uniform superposition of one sub-block}: \qquad \ |\tilde{s_2}\rangle=\frac{1}{\sqrt{\tilde{b}}}\sum^{\mbox{\tiny{$\tilde{b}$ items}}}_{\stackrel{\mbox{\tiny{one}}}{\mbox{\tiny{sub-block}}}}|x\rangle, \\ \nonumber \\ 
	\mbox{New local inversion}: \qquad \qquad \qquad \qquad \qquad \quad \ \
I_{\tilde{s_2}} = I-2|\tilde{s}_2\rangle \langle \tilde{s}_2|, \\
	\mbox{New local iteration}: \qquad \qquad \qquad \quad \ 
\tilde{G_2} = -I_{\tilde{s}_2}I_t, \qquad \sin^2\tilde{\theta}_2=\frac{1}{\tilde{b}}.	\label{newang2}
\end{eqnarray}
Written in these notations, the initial state of new database (\ref{ini}) is equivalent to a partially searched one with $\frac{\alpha\left(K\right)}{2}\sqrt{\tilde{N}}$ new global queries, i.e.
\begin{eqnarray}
	|d\rangle=G_1G_2^{j_2}G_1^{j_1}|s_1\rangle =  \tilde{G_1}^{\frac{\alpha\left(K\right)}{2}\sqrt{\tilde{N}}}|\tilde{s_1}\rangle \label{ininew}
\end{eqnarray}

Steps of sequential GRK can be written similarly to the first GRK using new notations (\ref{newnum})-(\ref{newang2}). The resultant state of target sub-block is
\begin{eqnarray}
|\tilde{d}\rangle \equiv \tilde{G_1}\tilde{G_2}^{\tilde{j_2}}\tilde{G_1}^{\tilde{j_1}}\left(\tilde{G_1}^{\frac{\alpha\left(K\right)}{2}\sqrt{\tilde{N}}}|\tilde{s_1}\rangle\right) = \sin \tilde{\omega} |t\rangle + \frac{\cos \tilde{\omega}}{\sqrt{\tilde{b}-1}} \sum^{\mbox{\tiny{$\tilde{b}-1$ items}}}_{\stackrel{x \neq t}{\mbox{\tiny{target sub-block}}}}|x\rangle.
\end{eqnarray}
Note that the vector in the parenthesis is $|d\rangle$ of (\ref{final}). We also have [similar to (\ref{vanish})]
\begin{eqnarray}
	\langle x|\tilde{d}\rangle = 0, \qquad \forall x \in \left\{\mbox{items of non-target sub-blocks}\right\}.
\end{eqnarray}
This yields cancellation condition relating $\tilde{j_1}$ and $\tilde{j_2}$, see \cite{Korepin}. 
We introduce parameters $\tilde{\eta}$ and $\tilde{\alpha}$ defined by
\begin{eqnarray}
	\tilde{j}_1=\left(\frac{\pi}{4}-\frac{\alpha(K)}{2} -\frac{\tilde{\eta}}{\sqrt{\tilde{K}}}\right) \sqrt{\tilde{N}} ,\qquad \tilde{j}_2=\tilde{\alpha} 
\sqrt{\tilde{b}}. \label{ssteps}
\end{eqnarray}
The algorithm is also optimized \cite{Korepin} in the large sub-block limit: $\tilde{b} \rightarrow \infty,\ \tilde{N}\equiv\tilde{K}\tilde{b} \rightarrow \infty$. In the limit, the angles (\ref{newang1}) and (\ref{newang2}) simplify
\begin{eqnarray}
	\tilde{\theta}_1= \frac{1}{\sqrt{\tilde{N}}}, \qquad	\tilde{\theta}_2=\frac{1}{\sqrt{\tilde{b}}}.
\end{eqnarray}
The minimum is achieved at
\begin{eqnarray}
	&\tilde{\eta} \left(\tilde{K}\right) = \frac{1}{2}\sqrt{\tilde{K}} \arctan \left(\frac{\sqrt{3\tilde{K}-4}}{\tilde{K}-2}\right) = \eta \left(\tilde{K}\right), \qquad    
	\tilde{\alpha} \left(\tilde{K}\right) = \frac{1}{2} \arccos \left(\frac{\tilde{K}-2}{2(\tilde{K}-1)}\right) =\alpha \left(\tilde{K}\right)&.
\end{eqnarray}
Similar to (\ref{omega}), we have in the large sub-block limit
\begin{eqnarray}
	\tilde{\omega}=\alpha(\tilde{K}).
\end{eqnarray}
As a result the number of queries of the sequential GRK is
\begin{eqnarray}
\bar{S}\left(K,\tilde{K}\right)\equiv \tilde{j}_1+\tilde{j}_2+1\stackrel{\tilde{b}\rightarrow\infty}{\longrightarrow}\left(\frac{\pi}{4}-\frac{\alpha(K)}{2}+\frac {\alpha \left(\tilde{K}\right)-\eta\left(\tilde{K}\right)}{\sqrt{\tilde{K}}} \right) \sqrt{\tilde{N}}. \label{no.seq}
\end{eqnarray}

In principle, sequential GRK's can be conducted successively until the smallest target sub-sub-block is found. Here arises a question on the efficiency of hierarchical partial search, i.e. whether or not is a sequence of GRK's works faster than a direct GRK partial search of the smallest sub-sub-blocks. 
As will be shown in the following section, direct GRK partial search makes less queries in the quantum case.


\section{Comparison of Hierarchical Partial Search with Direct Partial Search}

The partial search hierarchy forms a sequence of GRK's. It starts from searching for the largest target block and ends with searching for the smallest target sub-sub-block. On the other hand, it is also possible to partition the database directly into the smallest sub-sub-blocks and use a GRK to find the target sub-sub-block in one time. One question of significance is \textbf{whether the hierarchical search works faster than the direct search or not}. This question is of practical importance and the answer turns out to be \textit{negative}. We prove the statement by studying the first two successive GRK's in the hierarchy.

We have already derived the optimized number of queries of the first two GRK's in 
(\ref{no.par}) and (\ref{no.seq}), respectively. So that the total number of queries is the sum:
\begin{eqnarray}
	&T\left(K, \tilde{K}\right) \equiv S\left(K\right) + \bar{S}\left(K,\tilde{K}\right) \nonumber \\
&=\left\{\frac{\pi}{4}+\left[\frac{\pi}{4}+\frac{1}{2}\alpha\left(K\right)-\eta\left(K\right)\right]\frac{1}{\sqrt{K}}+\left[\alpha\left(\tilde{K}\right)-\eta\left(\tilde{K}\right)\right]\frac{1}{\sqrt{K\tilde{K}}}\right\}\sqrt{N}. \label{totseq}
\end{eqnarray}
On the other hand, if the database is partitioned directly into $K\tilde{K}$ blocks, a direct GRK algorithm would require
\begin{eqnarray}
	S\left(K\tilde{K}\right) = \left[\frac{\pi}{4}+\frac{\alpha\left(K\tilde{K}\right)-\eta\left(K\tilde{K}\right)}{\sqrt{K\tilde{K}}}\right]\sqrt{N} \label{totdir}
\end{eqnarray}
queries instead. Let us compare $T(K,\tilde{K})$ and $S(K\tilde{K})$, assuming that both $K\geq2$ and $\tilde{K}\geq2$.

\subsection{Numerical Comparison of Query Numbers and Asymptotic Analysis}
Before giving the complete proof, we illustrate this fact by looking at a few concrete examples. Here in Table \ref{tab:table} we give a few numerical examples of query numbers $S(K\tilde{K})$ and $T(K, \tilde{K})$ as well as their difference, for a better understanding. It is clear that each $T-S$ is positive in the last column.
\begin{table}[htbp]
\caption{\label{tab:table} Numerical Examples of Query Numbers}
\begin{ruledtabular}
\begin{tabular}{ccccc}
 $K$ & $\tilde{K}$ & $S(K\tilde{K})/\sqrt{N}$ & $T(K, \tilde{K})/\sqrt{N}$ & $(T(K, \tilde{K})-S(K\tilde{K}))/\sqrt{N}$\\
\hline
2 & 2 & 0.61548 & 0.670379 & 0.054899\\
2 & 3 & 0.646015 & 0.695421 & 0.049406\\
3 & 2 & 0.646015 & 0.721158 & 0.075143\\
2 & 4 & 0.664521 & 0.71289 & 0.048369\\
4 & 2 & 0.664521 & 0.73929 & 0.074769\\
3 & 3 & 0.671394 & 0.741605 & 0.070211\\
\end{tabular}
\end{ruledtabular}
\end{table}

Independently, we also look at the case that number of blocks and sub-blocks both being large, i.e. $K\rightarrow\infty$, $\tilde{K}\rightarrow\infty$. Asymptotic forms of $\alpha(x)$ and $\eta(x)$ are obtained from (\ref{min}) as
\begin{eqnarray}
	\alpha(x) \sim \frac{\pi}{6}+\frac{1}{2\sqrt{3} x}+
 \frac{5\sqrt{3}}{(6x)^2}, \qquad
 \eta(x) \sim  \frac{\sqrt{3}}{2}+\frac{1}{2\sqrt{3} x} +
\frac{11\sqrt{3}}{90x^2},
\qquad   x\rightarrow \infty. \label{asym}
\end{eqnarray}
Then the query numbers (\ref{totseq}) and (\ref{totdir}) take asymptotic forms using (\ref{asym})
\begin{eqnarray}
	&S(K\tilde{K}) \sim \left\{\frac{\pi}{4}+\left[\frac{\pi}{6}-\frac{\sqrt{3}}{2}+\frac{1}{5\sqrt{3}(2K\tilde{K})^2}\right]\frac{1}{\sqrt{K\tilde{K}}}\right\}\sqrt{N} \\
	&T(K, \tilde{K}) \sim \left\{\frac{\pi}{4}+\left[\left(\frac{\pi}{3}-\frac{\sqrt{3}}{2}\right)-\frac{1}{4\sqrt{3}K}-\frac{19\sqrt{3}}{10(6K)^2}\right]\frac{1}{\sqrt{K}}+\left[\frac{\pi}{6}-\frac{\sqrt{3}}{2}+\frac{1}{5\sqrt{3}(2\tilde{K})^2}\right]\frac{1}{\sqrt{K\tilde{K}}}\right\}\sqrt{N}.
\end{eqnarray}
As for the difference (\ref{diff}) of query numbers, the ratio $K/\tilde{K}$ becomes relevant in determining the asymptotic behavior. There are 3 possibilities:\\ 
If $K/\tilde{K}\rightarrow0$, then $1/K$ is dominating, and
\begin{eqnarray}
	&T(K, \tilde{K})-S(K\tilde{K}) \sim \left[\left(\frac{\pi}{3}-\frac{\sqrt{3}}{2}\right)K^{-\frac{1}{2}}\right]\frac{1}{\sqrt{N}}. \label{case1}
\end{eqnarray}
If $K/\tilde{K}\rightarrow\infty$, then $1/\tilde{K}$ is dominating, and
\begin{eqnarray}
	&T(K, \tilde{K})-S(K\tilde{K}) \sim \left(\frac{1}{20\sqrt
	{3}}K^{-\frac{1}{2}}\tilde{K}^{-\frac{5}{2}}\right)\frac{1}{\sqrt{N}}. \label{case2}
\end{eqnarray}
If $K/\tilde{K}\rightarrow\mbox{finite number}$, then we have the same result as (\ref{case1}). In both the expressions (\ref{case1}) and (\ref{case2}) the coefficients of $1/\sqrt{N}$ are positive. Up to now we saw that $T>S$. Now let us formally prove as a theorem (\ref{theorem}) that $T>S$ in general, when $K\geq2$ and $\tilde{K}\geq2$.

\subsection{General Proof that $T(K, \tilde{K})>S(K\tilde{K})$}

Now we prove that $T\left(K, \tilde{K}\right)-S\left(K\tilde{K}\right)$ is always positive in the region $K,\ \tilde{K}\in\left[2,+\infty\right)$. In order to complete the proof we need the following two lemmas.

\textbf{Lemma 1}: 
\begin{eqnarray}
	\frac{\pi}{4}+\left(\frac{1}{2}\alpha-\eta\right)\left(x\right)>0, \qquad \forall x\in \left[2,+\infty\right) \label{pro1}
\end{eqnarray}
Proof:	\qquad $\forall x\in \left[2,+\infty\right)$ \\
	The derivative $\left[\frac{\pi}{4}+\left(\frac{1}{2}\alpha-\eta\right)\right]^{\prime}\left(x\right)=\frac{1}{4\sqrt{x}}f\left(x\right)$ with $f\left(x\right)\equiv\frac{3}{\sqrt{3x-4}}-\arctan\frac{\sqrt{3x-4}}{x-2}$. While $f^{\prime}\left(x\right)=\frac{-9x+8}{2x(x-1)(3x-4)^{\frac{3}{2}}}<0$, so that $f\left(x\right)$ monotonous decreasing. Further, since that  $f\left(2\right)=\frac{3}{\sqrt{2}}-\frac{\pi}{2}>0,\ f\left(x\right)\stackrel{x\rightarrow+\infty}{\longrightarrow}0$, then continuous function $f\left(x\right)>0$ in the region. [$f(x)$ is positive at one point $x=2$ and tends to zero at infinity. As a continuous and monotonous function, $f(x)$ can never become negative nor zero in the region.] Therefore $\left[\frac{\pi}{4}+\left(\frac{1}{2}\alpha-\eta\right)\right]^{\prime}\left(x\right)>0$, so that $\frac{\pi}{4}+\left(\frac{1}{2}\alpha-\eta\right)\left(x\right)$ is a monotonous increasing function of $x$. With $\frac{\pi}{4}+\left(\frac{1}{2}\alpha-\eta\right)\left(2\right)=\frac{3-2\sqrt{2}}{8}\pi>0$, we conclude that $\frac{\pi}{4}+\left(\frac{1}{2}\alpha-\eta\right)\left(2\right)>0$ in the region.

\textbf{Lemma 2}:
\begin{eqnarray}
	\left(\alpha-\eta\right)\left(x\right) \ \mbox{monotonous decreasing}, \qquad \forall x\in \left[2,+\infty\right) \label{pro2}
\end{eqnarray}
Proof: \qquad $\forall x\in \left[2,+\infty\right)$ \\
The derivative $\left(\alpha-\eta\right)^{\prime}\left(x\right)=\frac{1}{4\sqrt{x}}g\left(x\right)$ with $g\left(x\right)\equiv\frac{\sqrt{3x-4}}{x-1}-\arctan\frac{\sqrt{3x-4}}{x-2}$. While $g^{\prime}\left(x\right)=\frac{1}{x\left(x-1\right)^{2}\sqrt{3x-4}}>0$, so that $g\left(x\right)$ monotonous increasing. Further, since that $g\left(2\right)=\sqrt{2}-\frac{\pi}{2}<0,\ g\left(x\right)\stackrel{x\rightarrow+\infty}{\longrightarrow}0$, then continuous function $g\left(x\right)<0$ in the region. [$g(x)$ is negative at one point $x=2$ and tends to zero at infinity. As a continuous and monotonous function, $g(x)$ can never become positive nor zero in the region.] Therefore $\left(\alpha-\eta\right)^{\prime}\left(x\right)<0$, we conclude that $\left(\alpha-\eta\right)\left(x\right)$ is a monotonous decreasing function of $x$ in the region.

Having proved these two lemmas, we look at the structure of $T\left(K, \tilde{K}\right)-S\left(K\tilde{K}\right)$ using (\ref{totseq}) and (\ref{totdir}): 
\begin{eqnarray}
	&T\left(K, \tilde{K}\right)-S\left(K\tilde{K}\right) \label{diff} \\
&=\left\{\left[\frac{\pi}{4}+\frac{1}{2}\alpha\left(K\right)-\eta\left(K\right)\right]\frac{1}{\sqrt{K}}+\left[\left(\alpha\left(\tilde{K}\right)-\eta\left(\tilde{K}\right)\right)-\left(\alpha\left(K\tilde{K}\right)-\eta\left(K\tilde{K}\right)\right)\right]\frac{1}{\sqrt{K\tilde{K}}}\right\}\sqrt{N}&. \nonumber
\end{eqnarray}
Making use of \textbf{Lemma 1} (\ref{pro1}), we see that the terms $\frac{\pi}{4}+\frac{1}{2}\alpha\left(K\right)-\eta\left(K\right)$ appearing in the first bracket of (\ref{diff}) is positive for $K\geq2$. Making use of \textbf{Lemma 2} (\ref{pro2}) and since $K\tilde{K}>\tilde{K}$, the monotony of $\alpha-\eta$ ensures that $\left(\alpha\left(\tilde{K}\right)-\eta\left(\tilde{K}\right)\right)>\left(\alpha\left(K\tilde{K}\right)-\eta\left(K\tilde{K}\right)\right)$. So that the second bracket of (\ref{diff}) is also positive for both $K\geq2$ and $\tilde{K}\geq2$. Therefore the whole brace of (\ref{diff}) is positive. As a consequence, we conclude our result in the following theorem

\textbf{Theorem}:
\begin{eqnarray}
	T\left(K, \tilde{K}\right)>S\left(K\tilde{K}\right), \qquad \forall K,\tilde{K} \in \left[2,\infty\right). \label{theorem}
\end{eqnarray}
i.e. Hierarchical partial search makes more queries to the oracle than direct partial search. \textit{Direct GRK partial search works faster}.

\subsection{Hierarchy with Many GRK's}

Theorem (\ref{theorem}) can be extended to the case of hierarchical search with an arbitrary number of GRK's. The direct GRK always works faster. We prove the statement as follows.

Consider a hierarchy with $m$ GRK's. Assume that $m\geq2$. We denote the whole operations $G_1G_2^{j_2}G_1^{j_1}$ of each GRK by one symbol and define an operator
\begin{eqnarray}
	{\cal G} \equiv G_1G_2^{j_2}G_1^{j_1}.
\end{eqnarray}
The hierarchical search works on the initial state $|s_1\rangle$ as
\begin{eqnarray}
	{\cal G}_m\ldots{\cal G}_3{\cal G}_2{\cal G}_1 |s_1\rangle,
\end{eqnarray}
where the sub-index denotes position of the GRK in the hierarchy [sequence]. The proof can be written formally in the following way. Define the total number of queries of the hierarchy
\begin{eqnarray}
	T(K_1, K_2,\ldots, K_m) \equiv S(K_1)+ \sum_{i=2}^{m} \bar{S}(K_{i-1},K_i). \label{def.T}
\end{eqnarray}
Here $K_i$ is number of "sub"-blocks in the $i^{\mbox{th}}$ partition of database. [We denoted $K_1$ and $K_2$ by $K$ and $\tilde{K}$ respectively in previous sections.] $S(K_1)$ is number of queries of the first GRK, and $\bar{S}(K_{i-1},K_1)$ that of the $i^{\mbox{th}}$ GRK in the hierarchy. Note that $S$ and $\bar{S}$ are not of the same function form. $S$ takes the form corresponding to a direct GRK (\ref{no.par}):
\begin{eqnarray}
	S(K_1)=\left(\frac{\pi}{4}+\frac{\alpha(K_1)-\eta(K_1)}{\sqrt{K_1}}\right)\sqrt{N}. \label{S_1}
\end{eqnarray}
While $\bar{S}$ takes a form of sequential GRK similar to (\ref{no.seq}):
\begin{eqnarray}
	\bar{S}(K_{i-1}, K_i)
	=\left(\frac{\pi}{4}-\frac{\alpha(K_{i-1})}{2}+\frac{\alpha(K_i)-\eta(K_i)}{\sqrt{K_i}}\right)\frac{\sqrt{N}}{\sqrt{\prod_{j=1}^{i-1}K_j}}, \qquad i\geq 2. \label{S_i}
\end{eqnarray}
[We denoted $S(K_1)$ and $\bar{S}(K_1,K_2)$ by $S(K)$ and $\bar{S}(K,\tilde{K})$ respectively in previous sections.] Let us substitute these expressions into (\ref{def.T}):
\begin{eqnarray}
	T(K_1, K_2,\ldots, K_m) = \left\{\frac{\pi}{4}+\sum_{i=1}^{m-1} \frac{\frac{\pi}{4}+\frac{1}{2}\alpha(K_i)-\eta(K_i)}{\sqrt{\prod_{j=1}^{i}K_j}}+\frac{\alpha(K_m)-\eta(K_m)}{\sqrt{\prod_{i=1}^{m}K_i}}\right\}\sqrt{N}. \label{T}
\end{eqnarray}
On the other hand, if we partition the database directly into the smallest sub-blocks, then the number of these sub-blocks would be $\prod_{i=1}^{m}K_i$. A direct GRK will locate the smallest target sub-block. This would require
\begin{eqnarray}
	S\left(\prod_{i=1}^{m}K_i\right)=\left\{\frac{\pi}{4}+\frac{{\alpha(\prod_{i=1}^{m}K_i)}-\eta(\prod_{i=1}^{m}K_i)}{\sqrt{\prod_{i=1}^{m}K_i}}\right\}\sqrt{N} \label{S}
\end{eqnarray}
queries to the oracle. Therefore the difference of (\ref{T}) and (\ref{S}) is
\begin{eqnarray}
	T\left(K_1, K_2,\ldots, K_m\right)-S\left(\prod_{i=1}^{m}K_i\right) \label{T-S} \qquad \qquad \qquad
	\qquad \qquad \\
=\left\{\left(\sum_{i=1}^{m-1}\frac{\frac{\pi}{4}+\frac{1}{2}\alpha(K_i)-\eta(K_i)}{\sqrt{\prod_{j=1}^{i}K_j}}\right)+\frac{\left[\alpha(K_m)-\eta(K_m)\right]-\left[\alpha(\prod_{i=1}^{m}K_i)-\eta(\prod_{i=1}^{m}K_i)\right]}{\sqrt{\prod_{i=1}^{m}K_i}}\right\}\sqrt{N}. \nonumber
\end{eqnarray}
We will show that this expression is always positive when each $K_i \geq 2$. Using \textbf{Lemma 1} (\ref{pro1}), we see that each term under the summation of (\ref{T-S}) is positive. Using \textbf{Lemma 2} (\ref{pro2}), $\alpha-\eta$ is a monotonous decreasing function. Note that product of all $K_i$'s is larger than $K_m$, we see that the remaining term in the brace of (\ref{T-S}) is also positive. Consequently, we conclude our result in the following corollary.

\textbf{Corollary}: 
\begin{eqnarray}
	T\left(K_1, K_2,\ldots, K_m\right)>S\left(\prod_{i=1}^{m}K_i\right), \qquad \forall K_i\in [2, +\infty). \label{corollary}
\end{eqnarray}
i.e. Hierarchy of \textit{arbitrary} number of GRK's makes more queries to the oracle than a direct GRK. \textit{Direct GRK partial search always works faster}.


\section{Summary}

The present paper studied quantum search. Partial search algorithm is called GRK. We studied partial search hierarchy and compared it with direct partial search [GRK]. Consider database of $N$ items with a single target item [target item also called marked item or solution]. The database is partitioned into $K$ blocks, each block further partitioned into $\tilde{K}$ sub-blocks. Hierarchical search is: use GRK and sequential GRK to find the target block and target sub-block, respectively. Successive GRK's can be made if the database is further partitioned. Each sequential GRK in the hierarchy works faster than the previous one. However, the total number of queries to the oracle adds up. The main conclusion is that \textit{a partial search hierarchy works slower than a direct partial search}, see theorem (\ref{theorem}) and corollary (\ref{corollary}). For example, consider a database partitioned into 3 blocks. Each block is further partitioned into 3 sub-blocks, so totally there are 9 sub-blocks. One could first find the target block using GRK, then the target sub-block by a sequential GRK. Nevertheless, it is faster to run a GRK partial search directly over the 9 sub-blocks and finds the target sub-block once.

\textbf{Note}: Only the class of algorithms using the standard Grover oracle was considered in the paper. This means that if one has already built the main Grover algorithm experimentally, then we do not need any new hardware to run the GRK algorithm. Another advantage of using the same oracle $I_t$ as the main Grover algorithm is more subtle: We can use ancilla [additional or auxiliary] q-bits to label different partitions of the database into blocks of equal size $b=N/K$. Then we are able to run GRK algorithm simultaneously for different partitions. [See Appendix D for more details.] Later a user can measure the ancilla q-bits and choose his or her favorite partition, by that time the target block already will be found.


\begin{acknowledgments}

The work is supported by NSF Grant DMS-0503712.

\end{acknowledgments}

\appendix
	\section{\newline Differences of the Last Operation of GRK in Literatures}
	
	Different versions of the last operation in \textbf{Step 3} of GRK appeared in literatures \cite{jaik, CK, Korepin}. People have finalized [after steps 1 and 2] the state $|v\rangle \equiv G_2^{j_2}G_1^{j_1}|s_1\rangle$ with different operations $I_{s_1}$, $-I_tI_{s_1}$, or $G_1\equiv-I_{s_1}I_t$. Grover and Radhakrishnan \cite{jaik} used $I_{s_1}$. This makes one less query to the oracle but the amplitude of the target item is negative in the final state $I_{s_1}|v\rangle$. Paper \cite{Korepin} used $-I_tI_{s_1}$ but paper \cite{KV} used $G_1$. The last two version become the same in the large block limit. This means that final states $-I_tI_{s_1}|v\rangle$ and $G_1|v\rangle$ are equivalent [of the same form] when $b\rightarrow\infty$, though $I_{s_1}$ and $I_t$ do not commute in general. We choose $G_1$ in our paper because it uses the same Grover iteration.
	\section{\newline Ranges of Parameters $\alpha$ and $\eta$}
	
	We are going to specify ranges of parameters $\alpha$ and $\eta$ introduced in (\ref{para}). Because of the constraint (\ref{constraint}) relating the two parameters, it is sufficient to specify the range of $\alpha$. It was shown in \cite{Korepin} that amplitudes [of items in the database after GRK] depend on $\sin{\left(2j_2\theta_2\right)}\sim\sin\left(2\alpha\right)$ and $\cos{\left(2j_2\theta_2\right)}\sim\cos\left(2\alpha\right)$. So that it is sufficient to take values of $\alpha$ within one period: $\alpha\in[a, a+\pi]$, with $a$ some real number determined later. We are looking for the exact boundaries of $\alpha$ set by physical considerations. 

Query numbers (\ref{para}) are non-negative:
\begin{eqnarray}
	j_1=\left(\frac{\pi}{4}-\frac{\eta}{\sqrt{K}}\right)\sqrt{N}\geq0, \label{j1} \\
	j_2=\frac{\alpha}{\sqrt{K}}\sqrt{N} \label{j2}\geq0.
\end{eqnarray}
Total query number (\ref{number}) should be less than that of a full Grover search:
\begin{eqnarray}
	j_1+j_2=\left(\frac{\pi}{4}+\frac{\alpha-\eta}{\sqrt{K}}\right)\sqrt{N}\leq\frac{\pi}{4}\sqrt{N}. \label{j1+j2}
\end{eqnarray}
These three inequalities (\ref{j1}), (\ref{j2}) and (\ref{j1+j2}) yield that
\begin{eqnarray}
	0\leq\alpha\leq\eta\leq\frac{\pi}{4}\sqrt{K}. \label{ineq}
\end{eqnarray}
We use constraint (\ref{constraint}) to express $\eta$ as a function of $\alpha$
\begin{eqnarray}
	\eta(\alpha)=\frac{1}{2}\sqrt{K}\mbox{Arctan}\left(\frac{2\sqrt{K}\sin{2\alpha}}{K-4\sin^2{\alpha}}\right)
\end{eqnarray}
with function Arctan(x) multi-valued. But according to (\ref{ineq}), we have
\begin{eqnarray}
	0\leq\mbox{Arctan}\left(\frac{2\sqrt{K}\sin{2\alpha}}{K-4\sin^2{\alpha}}\right)\leq\frac{\pi}{2}.
\end{eqnarray}
Therefore we could take the principal branch $\arctan{(x)}$. Now inequality (\ref{ineq}) becomes
\begin{eqnarray}
	0\leq\alpha\leq\frac{1}{2}\sqrt{K}\arctan\left(\frac{2\sqrt{K}\sin{2\alpha}}{K-4\sin^2{\alpha}}\right)\leq\frac{\pi}{4}\sqrt{K}. \label{main}
\end{eqnarray}
This inequality determines range of $\alpha$.

The solution of (\ref{ineq}) can be written in the following form:
\begin{eqnarray}
	0\leq\alpha\leq\alpha_{B}(K). \label{B} 
\end{eqnarray}
Here the upper bound $\alpha_{B}(K)$ is a function of $K$. When $K=2$, $3$ or $4$, $\alpha_{B}(K)$ coincide with the singularities of $\eta(\alpha)$. [ $K-4\sin^2{\alpha}=0$ at these singularities.] When $K\geq5$, values of $\alpha_{B}(K)$ can be solved numerically. As $K$ increases, $\alpha_{B}(K)$ approaches a certain positive number $\alpha_B(\infty)$. This limit $\alpha_{B}(\infty)=0.947747\ldots$ is the solution of $\alpha=\sin{(2\alpha)}$. [Inequality $\alpha\leq\eta(\alpha)$ becomes $\alpha\leq\sin{(2\alpha)}$ as $K\rightarrow\infty$.] The value of $\alpha_{B}(K)$ always lies in between $\alpha_{B}(\infty)$ and $\frac{\pi}{2}$ when $K\geq5$. We list these results in Table \ref{tab:bound}.

\begin{table}[htbp]
\caption{\label{tab:bound} Upper Bound of $\alpha$}
\begin{ruledtabular}
\begin{tabular}{cccccccc}
 $K$ & 2 & 3 & 4 & 5 & 6 & 100 & $\infty$\\
\hline
$\alpha_{B}(K)$ & $\frac{\pi}{4}$ & $\frac{\pi}{3}$ & $\frac{\pi}{2}$ & 1.22683 & 1.15100 & 0.956221 & 0.947747\\
\end{tabular}
\end{ruledtabular}
\end{table}
	\section{\newline Minimization of the Total Number of Queries of GRK}

Here we give a proof that (\ref{min}) is the global minimum of $\alpha-\eta$ under constraint (\ref{constraint}). In Appendix B we used (\ref{constraint}) to express $\eta$ as a function of $\alpha$
\begin{eqnarray}
	\eta\left(\alpha\right)=\frac{\sqrt{K}}{2}\arctan\left(\frac{2\sqrt{K}\sin2\alpha}{K-4\sin^2\alpha}\right).
\end{eqnarray}
Now we define a function
\begin{eqnarray}
	f\left(\alpha\right)\equiv \alpha-\eta\left(\alpha\right)
\end{eqnarray}
which we want to minimize within the range $0\leq\alpha\leq\alpha_{B}(K)$. We first prove that (\ref{min}) is a local minimum of $f(\alpha)$.
\subsection{Case $K\geq3$}
The first derivative of $f(\alpha)$ is
\begin{eqnarray}
	f^{\prime}(\alpha)=\frac{16(K-1)\sin^4{\alpha}-4K^2\sin^2\alpha+K^2}{16(K-1)\sin^4{\alpha}-8K\sin^2{\alpha}-K^2}.
	\label{derivative}
\end{eqnarray}
It vanishes at (\ref{min}) with $\sin^2{\alpha}=\frac{K}{4(K-1)}$. We calculate next the second derivative
\begin{eqnarray}
	&f^{\prime\prime}(\alpha)=\frac{4K\sin{2\alpha}[4(K-1)(K-2)\cos^2{2\alpha}+16(K-1)\cos{2\alpha}+(K-2)^2(K+2)]}{[16(K-1)\sin^4{\alpha}-8K\sin^2{\alpha}-K^2]^2}. \label{2ndderi}
\end{eqnarray}
Note that the value of the denominator at (\ref{min}) is $\frac{K^6}{(K-1)^2}$, which is strictly positive as $K\geq3$. The numerator is also positive because both $\sin{2\alpha}$ and $\cos{2\alpha}$ are positive at (\ref{min}) with $K\geq3$. [See \cite{Korepin} for the range of $\alpha(K)$.] Therefore $f^{\prime}(\alpha)=0$ and $f^{\prime\prime}(\alpha)>0$ at the solution (\ref{min}), so that (\ref{min}) is a local minimum for $K\geq3$.
\subsection{Case $K=2$}
The case that $K=2$ is more subtle. Expression (\ref{min}) yields that $\alpha=\frac{\pi}{4}$ and $\eta=\frac{\pi}{2\sqrt{2}}$. However, both first (\ref{derivative}) and second (\ref{2ndderi}) derivatives of $\alpha-\eta(\alpha)$ vanish at this critical point. The third derivative is non-zero: $f^{\prime\prime\prime}(\alpha=\frac{\pi}{4})=-4$. So we expand function $\alpha-\eta(\alpha)$ about the critical point
\begin{eqnarray}
	\alpha-\eta(\alpha)|_{K=2} = -\frac{4}{3!}\left(\alpha-\frac{\pi}{4}\right)^3 + {\cal O}\left(\left(\alpha-\frac{\pi}{4}\right)^4\right). \label{exp}
\end{eqnarray}
We see that $\alpha=\frac{\pi}{4}$ is actually a saddle point due to the non-vanishing cubic term. The form (\ref{exp}) suggests that if $\alpha$ goes greater than $\frac{\pi}{4}$, value of function $\alpha-\eta(\alpha)$ could be further reduced than the value at the saddle point. However, $\alpha=\frac{\pi}{4}$ is a boundary set by physical considerations [see Table \ref{tab:bound}]. Definition of $\alpha$ and $\eta$ in (\ref{para}) involves query numbers $j_1$ and $j_2$, which are non-negative. Therefore $j_1\equiv\left(\frac{\pi}{4}-\frac{\eta}{\sqrt{2}}\right)\sqrt{N}\geq0$, i.e. $\eta\leq\frac{\pi}{2\sqrt{2}}$. Now we allow $\alpha$ to go beyond $\frac{\pi}{4}$ and write
\begin{eqnarray}
	\alpha=\frac{\pi}{4}+\delta, \qquad
	\eta=\frac{\pi}{2\sqrt{2}}+\epsilon.
\end{eqnarray}
Here $\delta$ and $\epsilon$ are infinitesimals, $\delta>0$. Then constraint (\ref{constraint}) requires that
\begin{eqnarray}
	\epsilon=\delta.
\end{eqnarray}
So that $\eta$ would be greater than the physically allowed maximal value $\frac{\pi}{2\sqrt{2}}$ and $j_1$ would be negative $j_1=-\frac{\delta}{\sqrt{2}}\sqrt{N}$. This analysis showed that $\alpha$ can never go beyond $\frac{\pi}{4}$ and function $\alpha-\eta(\alpha)$ is minimized at this boundary. Therefore, expression (\ref{min}) as a local minimum is also valid in the case that $K=2$.

Now we have proved that the critical point (\ref{min})
\begin{eqnarray}
	\alpha \left(K\right) = \frac{1}{2} \arccos \left(\frac{K-2}{2(K-1)}\right) \label{crit}
\end{eqnarray}
is a local minimum of $f(\alpha)$. Note that $f(\alpha)$ is analytical as $0\leq\alpha\leq\alpha_{B}(K)$ and there is no singularity in this range any more. Therefore we can show that this local minimum (\ref{crit}) is also global by comparing the value of $f(\alpha)$ at (\ref{crit}) with those at the boundaries. [We always have $0<\alpha(K)\leq\alpha_{B}(K)$ and equality holds only for $K=2$.] We list the comparison results for $K=2$, $3$ and $4$ in Table \ref{tab:comp}.

\begin{table}[htbp]
\caption{\label{tab:comp} Comparison of values of $f(\alpha)$ at different points}
\begin{ruledtabular}
\begin{tabular}{cccc}
 $K$ & $f(0)$ & $f\left(\alpha(K)\right)$ & $f\left(\alpha_{B}(K)\right)$ \\
\hline
2 & 0 & $\frac{\pi}{4}\left(1-\sqrt{2}\right)\approx-0.325323$ & $\frac{\pi}{4}\left(1-\sqrt{2}\right)\approx-0.325323$\\
3 & 0 & -0.337098 & -0.313152\\
4 & 0 & -0.339837 & 0\\
\end{tabular}
\end{ruledtabular}
\end{table}

When $K\geq5$, $f(0)=f(\alpha_{B}(K))=0$, while $f(\alpha(K))<0$. Therefore, we conclude that the critical point (\ref{min}) or (\ref{crit}) is always the global minimum.

	\section{Different Partitions of a Database}
	
	A data base of N items can be partitioned into blocks in different ways. For example, items in one block may have the first 3 bits of their addresses the same for one partition or the last 3 bits the same for another partition. For a database partitioned into $K$ blocks of equal size $b=N/K$, there are totally
\begin{eqnarray}
	P(N,K)=\frac{N!}{\left(b!\right)^KK!}
\end{eqnarray}
different ways of partition. We could use ancilla q-bits [also called additional or auxiliary q-bits] to label these partitions. As $N$ and $b$ both being large, we shall need
\begin{eqnarray}
	\log_2P(N,K) \sim N\log_2 K -\log_2 K! \label{ancilla}
\end{eqnarray}
ancilla q-bits. For example, if we have $N=4$ items and $K=2$ blocks, then the number of partitions is $P(4,2)=3$ and we shall need $\log_{2}3\approx2$ ancillas. In practice, The number (\ref{ancilla}) can be further reduced if we only label the partitions commonly used, not all partitions. Then we can run GRK simultaneously for those selected partitions. When a user measures ancilla q-bit in his/her favorite partition, the target block will already be found by that time.

\end{document}